\documentclass[aip, apl, 10pt, groupedaddress, twocolumn]{revtex4-1}
\usepackage[T1]{fontenc}
\usepackage{bm}
\usepackage{graphicx}
\usepackage{amsmath}
\usepackage{amssymb}
\usepackage{sidecap}
\usepackage{dcolumn}
\usepackage{color}
\usepackage{ifpdf}
\usepackage{sidecap}

\begin{document}
\title{\textbf{Strain controlled valley filtering in multi-terminal graphene structures}}

\author{S. P. Milovanovi\'{c}}\email{slavisa.milovanovic@uantwerpen.be}
\author{F. M. Peeters}\email{francois.peeters@uantwerpen.be}

\affiliation{Departement Fysica, Universiteit Antwerpen \\
Groenenborgerlaan 171, B-2020 Antwerpen, Belgium}
\begin{abstract}
Valley-polarized currents can be generated by local straining of multi-terminal graphene devices. The pseudo-magnetic field created by the deformation allows electrons from only one valley to transmit and a current of electrons from a single valley is generated at the opposite side of the locally strained region. We show that valley filtering is most effective with bumps of a certain height and width. Despite the fact that the highest contribution to the polarized current comes from electrons from the lowest sub-band, contributions of other sub-bands are not negligible and can significantly enhance the output current.
\end{abstract}


\date{Antwerp, \today}

\maketitle

Conduction and valence band of graphene\cite{rgr01} touch in six points referred to as Dirac points. However, only two of them, labeled as $\textbf{K}$ and $\textbf{K'}$, are inequivalent and are related by time-reversal symmetry. This new degree of freedom opens the possibility to use these two valleys to encode information. However, the applicability of "valleytronics" relies on the assumption that the valley-polarized current can be easily generated and controlled. So far, there have been a number of propositions for such a device relying on the usage of ferromagnetic stripes \cite{rvf1, rvf11, rvf12}, nano-constrictions \cite{rvf2}, line defects \cite{rvf3, rvf31, rvf32}, specific substrates \cite{rvf4, rvf41}, etc. 

On the other hand, it was previosuly shown that graphene can sustain a large amount of strain. Due to its strong covalent $sp^2-$bonds, graphene can stretch up to 25$\%$ of its original size without breaking\cite{rstr01}. Mechanical deformations can lead to the generation of pseudo-magnetic fields (PMFs) that exceed \cite{rbb1} 300 T. Triaxial \cite{rts1, rts2} strain and strain generated by bending \cite{rbg1, rbg2} graphene give rise to a quasi-homogeneous PMF. Moreover, imperfections of the substrate can lead to the formation of bubbles and balloons that generate inhomogeneous PMFs \cite{rbb1, rbb2}. In the case of suspended devices a scanning tunnelling microscopy (STM) probe tip can be used to locally deform graphene membranes \cite{rsus1}. It was shown that the PMF generated in this manner produces regions with opposite sign of the pseudo-magnetic field\cite{rts2, rsus1, rgb1}.
\begin{figure}[htbp]
\begin{center}
\includegraphics[width=8.5cm]{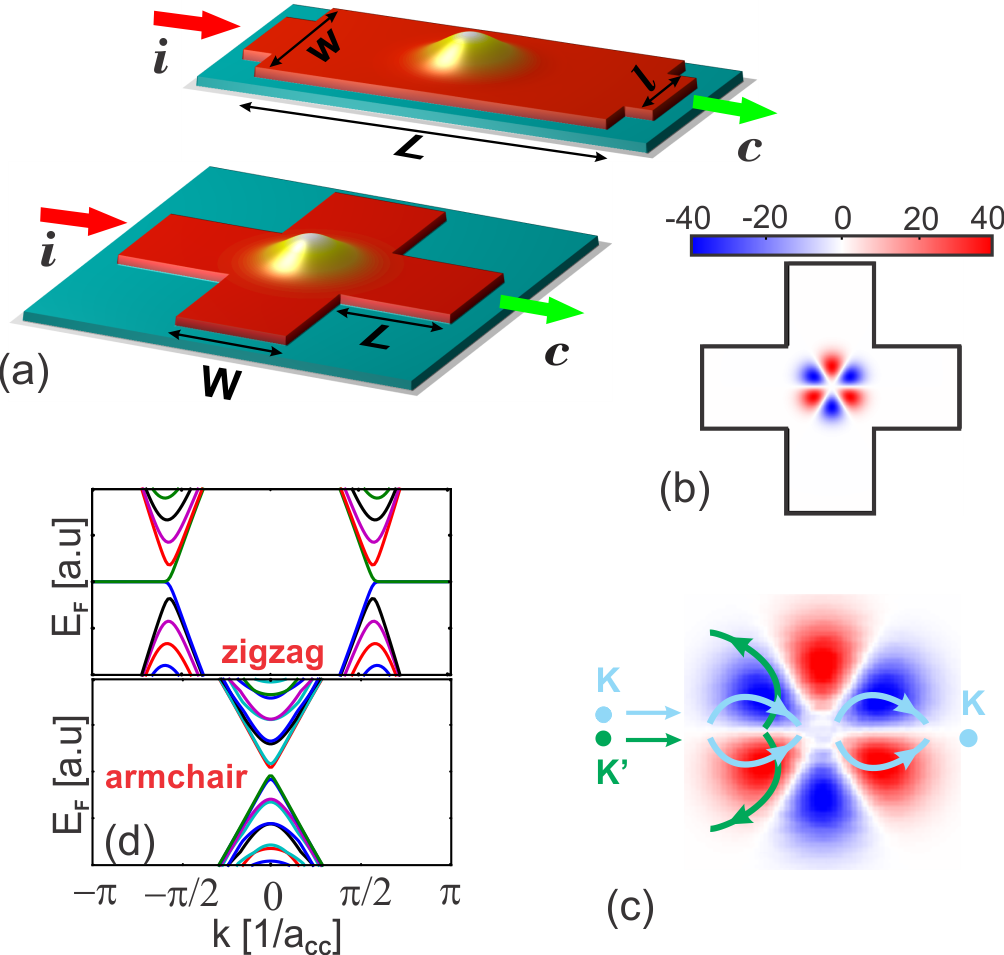}
\caption{(a) Schematic drawing of a two-terminal and a four-terminal device with a bump. (b) Profile of the pseudo-magnetic field induced by the bump for electrons residing in $\mathbf{K}$ valley. Calculations are made for a four-terminal device with $L = W = 100$ nm, $h_0 = 10$ nm, and $\sigma = 20$ nm. The magnetic field i.e. the color scale is given in Tesla. (c) Cartoon drawing of the flow of electrons from different valleys through the bump. (d) Energy dispersion relation for zigzag/armchair nanoribbon. Energy is shown in arbitrary units.}
\label{f1}
\end{center}
\end{figure}

The generated PMF has the opposite direction for electrons originating from different valleys. This property can be used in order to separate electrons from the two valleys and obtain a valley-polarized current \cite{rsett0, rsett1} - a prerequisite for valleytronics. Recently, Carrillo-Bastos \textit{et al.} showed that by using a Gaussian fold one is able to spatially separate electrons from opposite valleys \cite{rgf1}. Cavalcante \textit{et al.} proposed a particular straining of a graphene nanoribbon to realize a valley polarized current \cite{rgb2}. These proposals have the disadvantage that the valley polarization is difficult to tune and that actual devices are difficult to realize. In this Letter we show that straining graphene locally into a Gaussian bump by e.g. an AFM (atomic-force microscopy) tip, it is possible to obtain highly tunable polarized valley currents in two realistic device settings, shown in Fig. \ref{f1}(a): a two-probe nanoribbon and a Hall bar. 

Calculations are performed within the tight-binding model using standard nearest-neighbor Hamiltonian. Stretching graphene results in changes of the bond length between neighbouring atoms in its lattice. This change results in a modification of the hopping energy given by:
\begin{equation}
\label{e1}
t_{ij} = t_0 e^{-\beta(d_{ij}/a_0 - 1)},
\end{equation}
where $t_0 = -2.8$ eV, $\beta = 3.37$ is the strained hopping energy modulation factor, $a_0=0.142$ nm is the length of the unstrained $C-C$ bond, and $d_{ij}$ is the length of the strained bond between atoms $i$ and $j$. The change of hopping energy is equivalent to the generation of a magnetic vector potential \cite{rvp1}.
We consider a deformation that has a Gaussian profile given by,
\begin{equation}
\label{e4}
z(x, y) = h_0 e^{-(\vec{\mathbf{r}} - \vec{\mathbf{r_0}})^2/(2\sigma^2)},
\end{equation}
with $\vec{\mathbf{r}} = (x, y)$, and $\vec{\mathbf{r_0}} = (x_0, y_0)$ is the center of the bump. Applying the deformation given by Eq. \eqref{e4} we obtain the induced PMF \cite{rgb1} for electrons residing in the $\mathbf{K}$ valley as shown in Fig. \ref{f1}(b).

The profile of PMF for electrons from $\mathbf{K'}$ valley resembles the one from Fig. \ref{f1}(b) where the positive and negative regions are switched. This fact can be used to construct a valley filter where the idea is sketched in Fig. \ref{f1}(c). Electrons injected from different valleys feel the PMF of opposite sign and thus flow in opposite directions. Consequently, only electrons from one valley ($\mathbf{K}$) are able to transmit through the bump while electrons from the other valley ($\mathbf{K'}$) are immediately reflected. Hence, at the opposite side of the bump a valley-polarized current is generated. Of course, if the injector is at opposite side of the bump the situation is reversed. Now, due to the Lorenz force, electrons from $\mathbf{K'}$ valley pass through the bump while electrons from $\mathbf{K}$ valley are reflected. We test the valley filtering potential of our devices in the following way. Electrons are injected from lead $i$ (i.e. injector) and those electrons that pass the bump are collected at lead $c$ (i.e. collector). Pybinding \cite{rpb1}, a package for numerical tight-binding calculations, is used together with the Kwant package \cite{rkw1} to calculate the transmission probability between the two leads. In a graphene ribbon with zigzag edges the two valleys are well separated in momentum space and one can clearly distuinguish them, as shown in Fig. \ref{f1}(d). On the other hand, for a ribbon with armchair edges this is not the case. Thus, by choosing zigzag boundaries for injector/collector valley-polarized transmission $T_{\mathbf{K( K')}}$ can be easily computed from the scattering matrix\cite{rkw1} by separating those elements that belong to modes from opposite valleys. 
\begin{figure}[htbp]
\begin{center}
\includegraphics[width=8.5cm]{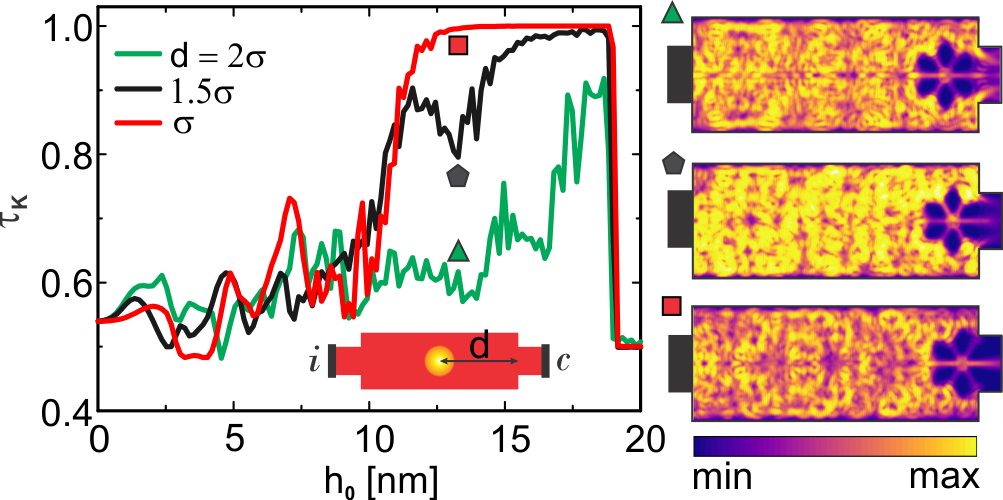}
\caption{Valley polarization $\tau_{\mathbf{K}}$ for two-terminal system shown in the inset versus the height of the bump for three different positions of the bump in $x-$direction. Other parameters are given in the text. Right panel shows current intensity plots for selected points indicated in the left figure.}
\label{f4}
\end{center}
\end{figure}

First, we test the valley filtering capabilities of a two-terminal device with $250 \times 100$ nm central region and two $50$-nm-wide leads attached to it. Fig. \ref{f4} shows the valley polarization of the output current, defined as $\tau_\mathbf{K} = T_\mathbf{K}/(T_\mathbf{K} + T_\mathbf{K'})$, versus the height of the bump. Calculations are performed using $E_F = 0.2$ eV and $\sigma = 15$ nm. We also show the influence of the position of the Gaussian bump on the valley polarization coefficient $\tau_{\mathbf{K}}$ by placing the bump at $d =  2\sigma, 1.5\sigma, \sigma$ away from the collector. The figure shows that the best result is obtained when the bump is placed closest to the collector and $\tau_{\mathbf{K}} \approx 1$ is achieved for $13$ nm $ < h_0 < 19$ nm. In the right panel we plot current intensities for a few points marked with symbols. The plot marked with red square shows the case of highest polarization. One can notice that the bump blocks the flow of electrons towards the collector. This is not surprising having in mind that high PMF formed inside the bump  reflects electrons. However, one can also notice a stream of electrons that flows along zero-PMF lines and into the collector. Since the output current is highly polarized we conclude that the stream consists entirely of electrons from $\mathbf{K}$ valley (see also Fig. \ref{f1}(c)). Hence, the bump efficiently valley filter injected current. If the center of the bump is moved away from the collector the valley polarization degrades. From the right panel of Fig. \ref{f4} it is clear that the current cannot penetrate inside the bump except along zero-PMF lines. However, when the bump is placed farther from the collector current can also flow around the bump and along the edges of the sample. Thus, the lower polarization is a consequence of leakage of non-polarized current into the collector around the edges of the bump. Furthermore, one can notice that for $h_0 = 0$ the polarization $\tau_{\mathbf{K}}$ is not 0.5 but slightly higher. The reason is the fully polarized zero-energy sub-band (see Fig. \ref{f1}(d)). While all other sub-bands are valley non-polarized this is not the case with this mode. Hence, when there is a low number of occupied sub-bands this mode introduces non-negligible imbalance to valley polarization even when there is no bump.

\begin{figure}[bp]
\begin{center}
\includegraphics[width=7.15cm]{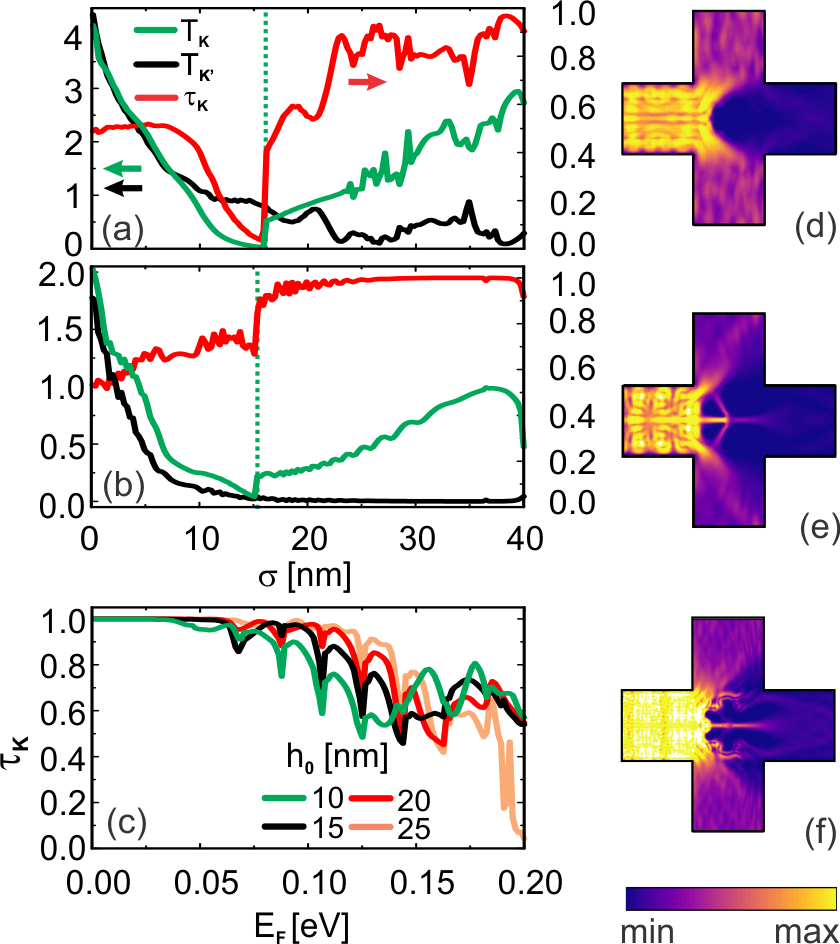}
\caption{ Probabilities $T_\mathbf{K}$, $T_\mathbf{K'}$, and  $\tau_\mathbf{K}$ versus the width of the Gaussian bump using $h_0 =  20$ nm and (a) $E_F = 0.2$ eV, (b) $E_F = 0.1$ eV. (c) Dependence of $\tau_\mathbf{K}$ on the Fermi energy for different values of $h_0$ given in the inset and $\sigma = 20$ nm. The current intensity plot for (d) $\sigma = 10$ nm, (e) $\sigma = 15$ nm with $E_F = 0.1$ eV, and (f) for $\sigma = 15$ nm with $E_F = 0.2$ eV.}
\label{f2}
\end{center}
\end{figure}

To improve the efficiency of our valley filter we need to eliminate leakage of non-polarized current into the collector. One way to do this is by adding additional leads that will collect this current. Therefore, we propose to use a four-terminal structure (Fig. \ref{f1}(a)) with $W = L = 100$ nm. In Figs. \ref{f2}(a) and (b) we plot the change of the transmission probabilities $T_\mathbf{K}$ and $T_\mathbf{K'}$ with the width of the bump, $\sigma$. Calculations are performed for two values of the Fermi energy, $E_F = 0.2$ eV and $E_F = 0.1$ eV, using $h_0 = 20$ nm. Both figures show similar behavior and one can distinguish two regimes - left and right from the dotted vertical line. In the former one, an increase of the width results in an overall decrease of the transmission probability. This is not surprising having in mind that the bump acts as a scatterer that blocks current flow towards the collector. Current intensity for one point in this regime is shown in Fig. \ref{f2}(d). We see that the electrons are unable to penetrate through the bump since high PMF formed inside the bump reflects them back. However, one can also notice that the stream of polarized electrons, observed in Fig. \ref{f4} along zero-PMF lines, does not appear. Hence, valley filtering is not possible in this regime. In the later one, we notice that an increase of $\sigma$ is followed by an increase of $T_\mathbf{K}$ while $T_\mathbf{K'}$ stays low, i.e. only $\mathbf{K}$ electrons flow into the collector. In this regime our device acts as a valley filter. We know that an increase of the width of the bump results in a decrease of the PMF (for constant $h_0$). Thus, in order to have valley filtering the PMF has to decrease enough to allow penetration of electrons through the bump and along the zero-PMF lines. This is shown in Fig. \ref{f2}(e) where we plot current intensity for $E_F = 0.1$ eV and $\sigma = 20$ nm.

Figs. \ref{f2}(a) and (b) show transmission probabilities for two systems that differ only in Fermi energy. However, polarization is higher in the case of lower Fermi energy. Therefore, in Fig. \ref{f2}(c) we plot the dependence of $\tau_\mathbf{K}$ on the Fermi energy for four values of $h_0$ given in the inset. Notice that at low Fermi energies $\tau_{\mathbf{K}}$ is unity for all values of $h_0$ which is due to the fact that at low energies only one sub-band is occupied and the injected current is already valley-polarized\cite{rvf2}. At higher energies curves exhibit a decrease of the current polarization. Decay of $\tau_{\mathbf{K}}$ starts at different values of $E_F$ for different $h_0$ which is related to the value of PMF. However, one can spot oscillations in $\tau_{\mathbf{K}}$ whose positions do not depend on $h_0$. These oscillations tell us at which energies new sub-bands start contributing to electric transport. The reason behind the decay of $\tau_{\mathbf{K}}$ with increase of $E_F$ can be understood from Fig. \ref{f2}(f) where we plot the current intensity for $E_F = 0.2$ eV while leaving other parameters as in Fig. \ref{f2}(e). We see that now electrons are able to transmit into the collector by flowing around the bump. Note that an increase of Fermi energy results in an increase of the cyclotron radius which allows electrons to penetrate deeper inside the bump effectively shrinking its width and the non-polarized current leaks into the collector.

\begin{figure}[htbp]
\begin{center}
\includegraphics[width=8.5cm]{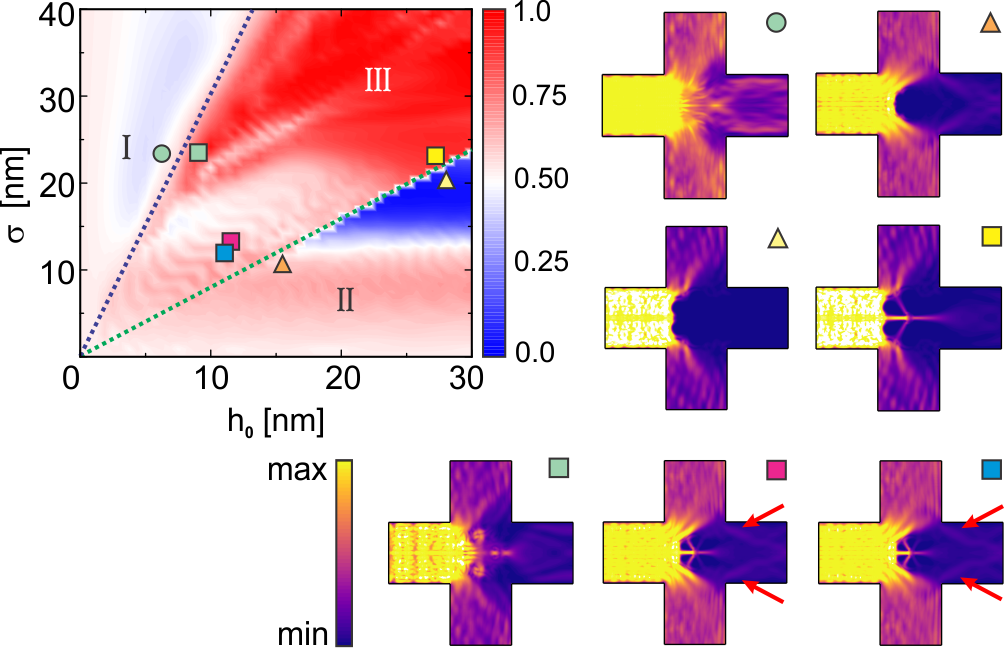}
\caption{Contour plot of valley polarization $\tau_{\mathbf{K}}$ versus the width and the height of the bump. Calculations are performed for $E_F = 0.15$ eV. In the right panel we show current intensity plots for different points from the contour plot marked by symbols.}
\label{f3}
\end{center}
\end{figure}

Contour plot of $\tau_{\mathbf{K}}$ versus the width and the height of the bump is shown in Fig. \ref{f3}. One can distinguish three different regimes. In regimes $\mathrm{I}$ and $\mathrm{II}$ valley filtering does not occur. Regime $\mathrm{I}$ corresponds to bumps with large $\sigma$ and small $h_0$ which generate low PMF that is unable to efficiently valley filter electrons. In the right panel of the same figure we plot current intensity for one point in this regime marked with green circle. Notice that the injected current beam transmits through the bump without filtering. We can conclude that by comparing this plot with the plot from Fig. \ref{f2}(e) which shows streams of valley-polarized electrons that flow only along zero PMF regions. In the later figure this is not the case since the current is well spread over the bump.

Regime $\mathrm{II}$ corresponds to bumps with small $\sigma$ and large $h_0$. The PMF generated with this combination of $\sigma / h_0$ is extremely high, hence, the bump repels all electrons, as seen from the current intensity plots marked with triangles. In this regime we can distinguish one "sub-regime" which appears for those values of $\sigma$ that are sufficiently large to completely block the collector for electrons from the $\mathbf{K}$ valley, as seen from current intensity plot marked with yellow triangle. High $\tau_{\mathbf{K'}}$ is a consequence of the fact that a very small percentage of electrons from the $\mathbf{K'}$ valley is still able to transmit into the collector by following the edges of the bump. Nonetheless, total current in the collector is on average 10 times smaller compered to the current outside this "sub-regime".

Regime $III$ is the regime where valley filtering is maximal. We calculated polarization plots for a few systems  with different terminal widths and different Fermi energies and found that valley filtering occurs for:
\begin{equation}
\label{e5}
c_1 < \sigma / h_0 < c_2,
\end{equation}
with dimensionless constant $c_1 \approx 0.8$ (green curve in Fig. \ref{f3}). The same value is extracted from results shown in Fig. \ref{f4} for a two-terminal device. Parameter $c_2$ is harder to extract due to the rather smooth transition between regimes $\mathrm{I}$ and $\mathrm{III}$. Furthermore, our study suggests that the upper limit depends on the value of the Fermi energy and width of the collector. This is not surprising since both parameters determine the amount of non-polarized current that leaks into the collector. Nevertheless, our results show that valley filtering occurs only in a specific range of $\sigma / h_0$. One can argue by observing Fig. \ref{f3} that Eq. \eqref{e5} is not a sufficient condition for the valley filtering regime and an additional restriction  $\sigma > \sigma_0$ needs to be imposed since $\tau_{\mathbf{K}}$ approaches unity only for larger values of $\sigma$. However, this is not completely true as it can be seen from current intensity plots shown in Fig. \ref{f3} for points in this regime. All four plots show streams of valley-polarized electrons flowing through the bump and the lower $\tau_{\mathbf{K}}$ is a consequence of the small size of the bump which allows electrons to flow around the bump into the collector (red arrows in Fig. \ref{f3}). 

We should mention that, in general, the transmission $T = T_{\mathbf{K}} + T_{\mathbf{K'}}$ is low. In order to have an efficient valley filter one has to have a large bump which reflects most of the injected electrons, as seen in Fig. \ref{f2}(e). 10$\%$ of electrons transmit through the bump. Our study shows that the highest contribution to $T$ comes from electrons from the lowest energy band which in the case of $E_F = 0.1$ eV contributes more than 95$\%$ to the output current. However, in the case of $E_F = 0.2$ eV transmission $T_\mathbf{K}$ is much larger than unity which means that contributions of other bands are of importance.

In conclusion, valley filtering capabilities of different multi-terminal graphene devices with a Gaussian bump were investigated. Our study showed that in order to construct an efficient valley filter with a Gaussian bump one needs to satisfy the following requirements. Valley filtering is only possible with bumps that satisfy  $0.8 < \sigma/h_0 < c_2$, where the upper limit depends on the Fermi energy and width of the terminal. Furthermore, the polarization of output current decreases with increase of the Fermi energy. This was explained by the fact that for higher Fermi energies electrons are able to penetrate deeper inside the bump, effectively shrinking its width, which allows current flow around the bump and into the collector. We examined the dependence of polarization on the position of the bump and concluded that more efficient valley filter requires a bump placed closer to the collector.

This work was supported by the Flemish Science Foundation (FWO-Vl) and the European Science Foundation (ESF) under the EUROCORES Program EuroGRAPHENE within the project CONGRAN.

\begin{thebibliography}{99}
%
%
%
%
\bibitem{rgr01} K. S. Novoselov, A. K. Geim, S. V. Morozov, D. Jiang, Y. Zhang, S. V. Dubonos, I. V. Grigorieva, and A. A. Firsov, Science \textbf{306}, 666 (2004).
%
\bibitem{rvf1} T. Fujita, M. B. A. Jalil, and S. G. Tan, Appl. Phys. Lett. \textbf{97}, 043508 (2010).
%
\bibitem{rvf11} Z. Wu, F. Zhai, F. M. Peeters, H. Q. Xu, and K. Chang, Phys. Rev. Lett. \textbf{106}, 176802 (2011).
%
\bibitem{rvf12} F. Zhai, Y. Ma, and Y.-T. Zhang, J. Phys.: Condens. Matter \textbf{23}, 385302 (2011).
%
\bibitem{rvf2} A. Rycerz, J. Tworzyd\l{}o, and C. W. J. Beenakker, Nat. Phys. \textbf{3}, 172 (2007). 
%
\bibitem{rvf3} D. Gunlycke and C. T. White, Phys. Rev. Lett. \textbf{106}, 136806 (2011).
%
\bibitem{rvf31} J.-H. Chen, G. Aut\`{e}s, N. Alem, F. Gargiulo, A. Gautam, M. Linck, C. Kisielowski, O. V. Yazyev, S. G. Louie, and A. Zettl, Phys. Rev. B \textbf{89}, 121407(R) (2014).
%
\bibitem{rvf32} V. H. Nguyen, P. Dollfus, and J.-C. Charlier, arXiv:1605.07777 (2016).
%
\bibitem{rvf4} D. Moldovan, M. Ramezani Masir, L. Covaci, and F. M. Peeters, Phys. Rev. B \textbf{86}, 115431 (2012).
%
\bibitem{rvf41} M. M. Gruji\'{c}, M. \v{Z}. Tadi\'{c}, and F. M. Peeters, Phys. Rev. Lett. \textbf{113}, 046601 (2014).
%
\bibitem{rstr01} C. Lee, X. Wei, J. W. Kysar, and James Hone, Science \textbf{321}, 385 (2008).
%
\bibitem{rbb1} N. Levy, S. A. Burke, K. L. Meaker, M. Panlasigui, A. Zettl, F. Guinea, A. H. Castro Neto, and M. F. Crommie, Science \textbf{329}, 544 (2010).
%
\bibitem{rts1} F. Guinea, M. I. Katsnelson, and  A. K. Geim, Nat. Phys. \textbf{6}, 30 (2009).
%
\bibitem{rts2} M. Ramezani Masir, D. Moldovan, and F.M. Peeters, Solid State Comm. \textbf{175-176}, 76 (2013).
%
\bibitem{rbg1} F. Guinea, A. K. Geim, M. I. Katsnelson, and K. S. Novoselov, Phys. Rev. B \textbf{81}, 035408 (2010).
%
\bibitem{rbg2} T. Low and F. Guinea, Nano Lett. \textbf{10}, 3551 (2010).
%
\bibitem{rbb2} J. Zabel, R. R. Nair, A. Ott, T. Georgiou, A. K. Geim, K. S. Novoselov, and C. Casiraghi, Nano Lett. \textbf{12}, 617 (2012).
%
\bibitem{rsus1} N. N. Klimov, S. Jung, S. Zhu, T. Li, C. A. Wright, S. D. Solares, D. B. Newell, N. B. Zhitenev, and J. A. Stroscio, Science \textbf{336}, 1557 (2012).
%
\bibitem{rgb1} D. Moldovan, M. Ramezani Masir, and F. M. Peeters, Phys. Rev. B \textbf{88}, 035446 (2013).
%
\bibitem{rsett0} M. Settnes, Doctoral dissertation, DTU Nanotech (2015).
%
\bibitem{rsett1} M. Settnes, S. R. Power, M. Brandbyge, and A.-P. Jauho, arXiv:1608.04569 (2016).
%
\bibitem{rgf1} R. Carrillo-Bastos, C. Le\'{o}n, D. Faria, A. Latg\'{e}, Eva Y. Andrei, and N. Sandler, arXiv:1604.00732 (2016).
%
\bibitem{rgb2} L. S. Cavalcante, A. Chaves, D. R. da Costa, G. A. Farias, and F. M. Peeters, arXiv:1606.09226 (2016).
%
\bibitem{rpb1} D. Moldovan and F. M. Peeters, DOI: 10.5281/zenodo.56818 (2016).
%
\bibitem{rkw1} C. W. Groth, M. Wimmer, A. R. Akhmerov, and X. Waintal, New J. Phys. \textbf{16}, 063065 (2014).
%
\bibitem{rvp1} V. M. Pereira and A. H. Castro Neto, Phys. Rev. Lett. \textbf{103}, 046801 (2009).
%
\end{thebibliography}
\end{document}